\documentclass[12pt]{article}
\usepackage{times,parskip,apalike,epsf,epsfig,latexsym,rotating}
\begin{document}

\title{Power Euclidean metrics for covariance matrices with application 
to diffusion tensor imaging}

\author{Ian L.~Dryden$^{+,*}$, Xavier Pennec$^{\dagger}$ and Jean-Marc Peyrat$^{\dagger}$\\
($+$) University of Nottingham, UK,\\
($*$) University of South Carolina, USA\\
($\dagger$) INRIA, Sophia Antipolis, France.}

\date{}
\maketitle

\begin{abstract}
Various metrics for comparing diffusion tensors                          
have been recently proposed in the literature. We consider a broad
family of metrics which is indexed by a single power parameter. 
A likelihood-based procedure is developed for choosing the 
most appropriate metric from the family for a given dataset at hand. 
The approach is analogous to using the Box-Cox transformation that is  frequently 
investigated in regression analysis. The methodology is illustrated with a simulation 
study and an application to a real dataset of diffusion tensor images 
of canine hearts. 
\end{abstract} 


\section{Introduction}

In many applications it is of interest to consider statistical analysis of covariance matrices. 
In particular, typical tasks might be to estimate a mean covariance matrix, describe the 
structure of the variability in the matrices, 
and interpolate between covariance matrices. A key aspect of
such analysis is the choice of metric for comparing covariance matrices. 

In medical imaging 
a particular type of data structure commonly arises when analysing diffusion weighted images, 
called the diffusion tensor (Basser et al., 1994; Alexander, 2005). \nocite{Basseretal94,Alexander05} The diffusion tensor (DT) 
is simply a positive semi-definite symmetric matrix which
is related to the covariance matrix of the diffusion of water molecules at a particular 
voxel in the brain.
In diffusion tensor
imaging a number of different metrics have been proposed for comparing two tensors 
$S_1, S_2$, which are symmetric positive semi-definite $m \times m$ covariance matrices.
Metrics include the Euclidean metric,  Riemannian metric (Pennec et al., 2006),  
log-Euclidean metric (Arsigny et al., 2007), 
Cholesky metric (Wang et al., 2004), and the Procrustes size-and-shape metric 
(Dryden et al., 2009), among many others. In several of these papers it has been demonstrated how
the choice of metric makes a substantial difference to estimation and interpolation. 
However, for a particular application at hand, which is the preferred choice of metric?

 \nocite{Arsignyetal07,Drydenetal09}
\nocite{Pennecetal06,Wangetal04}

Statistical approaches to the analysis of DT  data include 
Schwartzman et al. (2010), Whitcher et al. (2007), Zhu et al. (2007, 2009), 
and, for example, hypothesis tests for group means and extensions 
of regression models are
considered. However, the choice of statistical model or metric on
the space of diffusion tensors is problem specific.
\nocite{Schwartzmanetal10,Whitcheretal07,Zhuetal07,Zhuetal09}

The aim of this paper is to consider a family of metrics indexed
by a single power parameter. We will develop likelihood-based methods for 
estimation of the power parameter, which will enable us to select the most appropriate 
metric from the family. 
The methodology is applied to simulated data and a real 
dataset of diffusion tensors from a sample of canine hearts. The procedure that is developed 
is directly analogous to the use of variance stabilizing Box-Cox transformations that are very commonly used in 
regression analysis in statistics (Box and Cox, 1964). \nocite{Boxcox64} 
The general idea is that a suitable metric for a problem will be based on a power transformation
that makes the data ``as Gaussian as possible". 

\section{Power Euclidean metrics}
\subsection{Metrics and the sample Fr\'{e}chet mean} 
Let $S = U \Lambda U^T$ be the usual spectral decomposition with $U \in O(m)$ an orthogonal matrix and 
 $\Lambda$ diagonal with strictly positive entries. Let $\log \Lambda$ be a diagonal matrix with 
logarithm of the diagonal elements of $\Lambda$ on the diagonal. The logarithm of $S$ is 
given by $\log S = U ( \log \Lambda ) U^T$ and likewise the exponential of the matrix $S$ is  
  $\exp S = U (\exp \Lambda) U^T$. The log-Euclidean metric (Arsigny et al., 2007) 
is given by 
\begin{equation}
d_L(S_1,S_2) = \|  \log(S_1) - \log(S_2)  \|  \; . \label{dL}
\end{equation}

A broad family of covariance matrix metrics is the set of power Euclidean metrics
\begin{equation}
d_{A}(S_1, S_2) =  
  \frac{1}{\alpha} \|  S_1^{\alpha}  - S_2^{\alpha}  \| ,  \label{dA}
\end{equation}
where $S^{\alpha} = U \Lambda^{\alpha} U^T$ (see Dryden et al., 2009). As $\alpha \to 0$ the metric approaches the 
log-Euclidean metric of (\ref{dL}). 
 We could consider any $\alpha \in \mathbf R$ depending on the situation, 
 and we use the log-Euclidean metric if $\alpha=0$.
Note that all these metrics are Euclidean metrics, with curvature of the space being zero. 
Tangent space co-ordinates are also therefore very simple. 
If $\alpha = 0$ the covariance matrices must be strictly positive definite, but 
if $\alpha \ne 0$ then we can compare positive semi-definite matrices.

Consider a random sample available $\{ S_1,\ldots, S_n \}$ from a 
population. The 
sample Fr\'{e}chet mean based on metric $d()$ is 
calculated by finding 
$$\hat \Sigma = \mathop{{\rm arg}\inf_{\Sigma}} \sum_{i=1}^n d( S_i, \Sigma)^2 . $$
In our case with all the power metrics, the manifold is isomorphic to a
convex subset of $R^p$, and the existence and uniqueness of the mean
  follows immediately   (Karcher, 1977; Le, 1995). 
\nocite{Karcher77,Le95}

The sample Fr{\'e}chet mean of the powered covariance matrices provides an
estimate of the population covariance matrix, and is given by
$$
\hat \Sigma_A = ( \hat \Delta_A )^{1/\alpha} , \; \; {\rm where} \; \; \;  
 \hat \Delta_A = \mathop{{\rm arg}\inf_{\Delta}}  \left\{  \sum_{i=1}^n \| S_i^{\alpha} - \Delta \|^2 \right\} 
= \frac{1}{n} \sum_{i=1}^n S_i^{\alpha}  .  
$$
For $\alpha > 0$ the estimators become more resistant to outliers for smaller $\alpha$, and for larger $\alpha$ the
the estimators become less resistant to outliers. For $\alpha < 0$ one is working with powers
of the inverse covariance matrix. 
The resulting fractional anisotropy measure using the power metric (\ref{dA}) is given by 
$$
FA(\alpha) = \left\{ \frac{m}{m-1} \sum_{i=1}^m (\lambda_i^{\alpha} - \overline { \lambda^{\alpha }} )^2 / 
\sum_{i=1}^m \lambda_i^{2 \alpha}  \right\}^{1/2} ,$$
and $\overline { \lambda^{\alpha } } = \frac{1}{m} \sum_{i=1}^m \lambda_i^{\alpha}$. 
A practical visualisation tool is to vary $\alpha$ in order for a neurologist 
to help interpret the white fibre tracts in the images (Dryden et al., 2009).

A question of interest is what $\alpha$ should we take in practice? We develop a criterion 
which involves choosing $\alpha$ to make the data as multivariate Gaussian as possible, as 
statistical inference should be more straightforward in such cases.
 Such a criterion is similar to the Box-Cox transformation used in
regression analysis, except in our case we are working with 
matrix powers of symmetric positive definite matrices.

\subsection{Likelihood}
First of all we write down the probability density      function for  $L_i = S_i^{\alpha}$, 
$i=1,\ldots,n$, which is assumed to be multivariate Gaussian in the $m(m+1)/2$
distinct elements of 
$L_i$. The density of $L_i$ is 
$$f_L( L_i ; \mu, \Sigma) = \frac{1}{(2 \pi )^{m(m+1)/4}} | \Sigma |^{-1/2} \exp \{ - \frac{1}{2 }  {\rm vech}(L_i - \mu)^T \Sigma^{-1} 
{\rm vech}(L_i - \mu)   \} ,\; \; \; \; \; i=1,\ldots,n,$$
where the parameters are $\mu$ (a $m(m+1)/2$ vector) and $\Sigma$ (a $m(m+1)/2 \times m(m+1)/2$ 
symmetric positive definite matrix). Note that ${\rm vech}(A)$ is a vector
of the $m(m+1)/2$ elements in the upper triangle of $A$.

Our data though are given as $S_i$, so we transform from $L$ to $S$ using 
$ S = (\alpha L)^{1/\alpha}$, to give the density of $S$.
The transformation can be carried out in three stages.
We first transform from $L$ to $(U,\frac{1}{\alpha}D^{\alpha})$, where $L = \frac{1}{\alpha}UD^{\alpha}U^T$ 
is the usual 
spectral decomposition with $U \in O(m)$ and $D = {\rm diag}(d_1,\ldots,d_m)$ 
is diagonal with positive, distinct diagonal values. 
We then transform from $\frac{1}{\alpha}D^{\alpha}$ to $D$ and then finally we transform to $S = 
UDU^T$.
Using Fang and Zhang (1990, p24) \nocite{Fangzhang90} the Jacobian of the first transformation is
given by 
$$ J_1 = {f_n(U)} \prod_{i=1}^{m-1} \prod_{j=i+1}^m  \frac{1}{\alpha}
( d_i^{\alpha} - d_j^\alpha) , $$
where $f_n(U) = 1/\prod_{i=1}^n | U_{(i)}|_+$ and $U_{(i)}$ is the 
first $i$-dimensional principal minor of $U$.
The second transformation has Jacobian 
$$ J_2 = \prod_{i=1}^m d_i^{\alpha-1} , $$
and finally the third transformation has Jacobian 
$$ J_3 = \frac{1}{f_n(U)} \prod_{i=1}^{m-1} \prod_{j=i+1}^m  \frac{1}
{ d_i - d_j} . $$
Putting these togther the density of $S$ given $\alpha$ is:
  $$f_S(S; \alpha, \mu, \Sigma)   = f_L( \frac{1}{\alpha}S^{\alpha} ; \mu, \Sigma)  
\prod_{i=1}^m d_i^{\alpha -1}  
 \prod_{i=1}^{m-1} \prod_{j=i+1}^m  \frac{d_i^{\alpha}-d_j^{\alpha}}{\alpha( d_i - d_j)}  .$$

Hence, given a random sample $S_1,\ldots,S_n$ of diffusion tensors 
the log-likelihood of the parameters $(\alpha,\mu, \Sigma)$ is 
given by 
\begin{equation}
 L(\alpha, \mu, \Sigma) = \sum_{i=1}^n \log f_S(S_i ; \alpha , \mu, \Sigma). \label{likedef}
\end{equation}

\subsection{Statistical Inference}
Maximum likelihood estimation of the parameters in the model is straightforward if
$\alpha$ is known, in which case $\mu$ and $\Sigma$ are given by 
the usual Gaussian m.l.e.s of the 
sample mean and sample covariance matrices of ${\rm vech}(L_i), i=1,\ldots,n$, 
where $n > m(m+1)/2$.   

Hence, a method for estimation is to evaluate the profile log-likelihood 
for a range of $\alpha$ (substituting in the maximum likelihood estimators 
(m.l.e.'s) of 
$\mu, \Sigma$ given $\alpha$). The Jacobian term is straightforward to evaluate if the eigenvalues of $S_i$ are sufficiently distinct. However, if 
some of them are close to each other 
then a Taylor series expansion is used for numerical 
evaluation. Also, a Taylor series expansion is used for $\alpha$ close
to zero.
In particular, writing $\mu = \lambda_b/\lambda_a -1$ then we have the Taylor Series expansion
\begin{eqnarray}
\frac{ \lambda_a^{\alpha} - \lambda_b^{\alpha} }
{\alpha ( \lambda_a          - \lambda_b         ) }
& = &  
\lambda_a^{\alpha-1} ( (1 + \mu)^\alpha - 1) / (\alpha \mu) \nonumber \\
& = & \lambda_a^{k-1}(1+\frac{1}{2}(\alpha-1)\mu + \frac{1}{6}(\alpha-1)(\alpha-2) \mu^2 +
 \frac{1}{24}(\alpha-1)(\alpha-2)(\alpha-3) \mu^3 +O(\mu^4)) \nonumber
\end{eqnarray}
which can be used for computation when the eigenvalues become close. Also, the Taylor series 
expansion in terms of $\alpha$ is 
$$\lambda_a^{\alpha-1} ( \log(1+\mu)/\mu+ \alpha \log(1+\mu)^2/(2\mu)+\alpha^2 \log(1+\mu)^3/(6\mu)+
\alpha^3\log(1+\mu)^4/(24\mu)+O(\alpha^4)) , $$
which is useful for small $|\alpha|$.

In order to obtain confidence intervals for $\alpha$ we 
use Wilks' Theorem which states that, for large samples, 
$$ -2 \log \Lambda \approx \chi_1^2  , $$
where $\Lambda$ is the likelihood ratio for testing $H_0: \alpha = \alpha_0$ 
versus $H_1: \alpha \ne \alpha_0$. In practice, an approximate 95\% 
confidence interval is obtained by the values of $\alpha$ which have 
log-likelihood within $2$ from the maximum.

\section{Applications}
\subsection{Simulation study}
We initially consider a simulation study with $m=3$ and 
${\rm vech}(X)$ are simulated from a
multivariate Gaussian 
$${\rm vech}(X) \sim N_6(\mu,\sigma^2 I) \; , $$
and then the symmetric matrix $X$ is 
transformed to the matrix $S = (\alpha X)^{1/\alpha}$. 
A random sample of $S$ matrices are simulated for $n_v$ voxels, with 
$n_s$ samples at each voxel. Each of the $n =  n_sn_v$ simulated 
tensors are independent and identically distributed. We consider an example 
simulation where the true mean matrix is 
$\mu = {\rm diag}(2,1,1)$, the true $\alpha = 0.3$ and $\sigma^2 = 0.02$ is
the noise variance. 

For each simulated sample of size $n$ the m.l.e. and 
confidence intervals for $\alpha$ are obtained, by 
evaluating the log-likelihood over a range of values for $\alpha$. 
In our example the range of values for evaluation 
is $[-0.1,0.7]$ in steps of $0.02$. 
The m.l.e. of $\alpha$ maximizes the log-likelihood (\ref{likedef}), 
and the confidence interval is taken as the range of $\alpha$ 
within $2$ of the maximum of the log-likelihood. 

From $1000$ Monte Carlo simulations we count the number of times that the 
true $\alpha$ lies inside the estimated $95\%$ confidence interval. 
The results are given in Table \ref{TAB1}. We see that the coverage is 
low for small samples, with the confidence intervals being conservative. 
However, coverage is very good for $n_s n_v \ge 20$ here.

\begin{center}
{\bf INSERT TABLE 1 ABOUT HERE}
\end{center}

\subsection{Canine Hearts}
The estimation of $\alpha$ and confidence intervals are also carried 
out for data on a subset of registered DT images of canine hearts. 
The dataset consists of nine registered hearts, and is described by 
Peyrat et al. (2007). \nocite{Peyrat:TMI:2007}
There are $n_s=9$  DT images which have been registered 
and then a mask is extracted. We take an equally spaced grid at intervals of 
$2$mm in the mask 
and consider a neighbourhood of voxels within a ball of radius $0.7$mm. We consider interior 
points where the sub-region contains at $n_v-1$ neighbourhood voxels in addition to the voxel itself 
(where $n_v \ge 15$).  
Hence, we have $n = n_s n_v$ voxels contributing to the likelihood 
of (\ref{likedef}), with a common $\mu, \Sigma$ and $\alpha$. The parameter 
$\alpha$ is estimated by maximum likelihood using the voxels in each local sub-region. 
Hence, we are able to estimate $\alpha$ in different regions of the heart. 

We consider analysis of the DT data using a global normalization (as the data have been 
measured at different temperatures) and also without normalization. 
The DTs have been normalised for each heart by taking an overall mean 
tensor (using Euclidean mean $\alpha = 1$) 
and then dividing by the norm of this average matrix.

\begin{figure}[htbp]
\begin{center}
\includegraphics[width=8cm,angle=270]{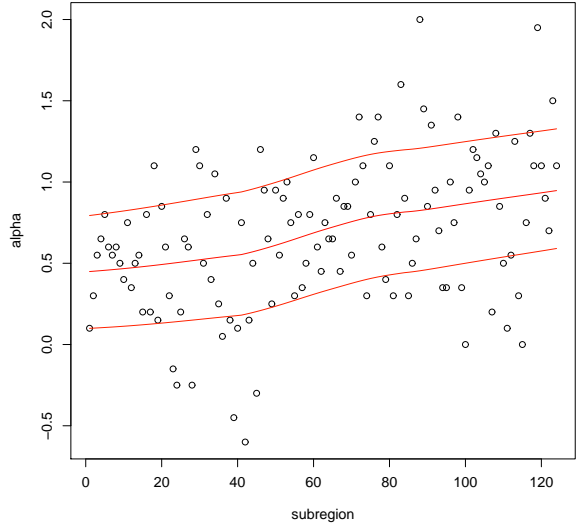}
\end{center}
\caption{The m.l.e. of $\alpha$ in each of the $124$ neighbourhoods. Each image has been 
normalised by dividing by the global norm of the average DT matrix. The location of the 
neighbourhoods proceeds from left of the mask (lower neighbourhood labels) to the right (higher numbered labels). 
A smoother estimate of the m.l.e. is given (middle line) together with smoothed upper and 
lower 95\% confidence limits for each neighbourhood.} \label{FIG1}
\end{figure}

\begin{figure}[htbp]
\begin{center}
\includegraphics[width=8cm]{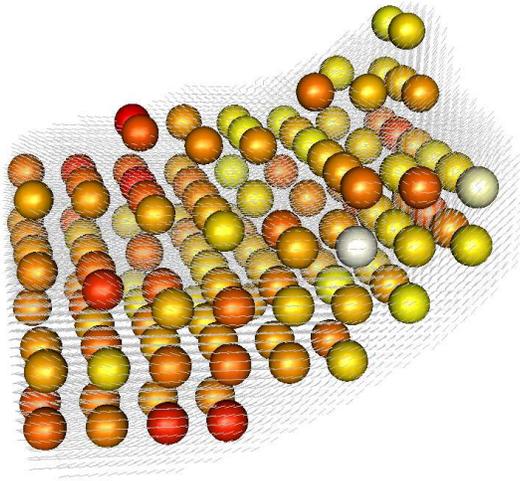}
\end{center}
\caption{The mask is shown with the principal axis of each DT for the first canine 
heart. The coloured spheres show the estimated $\alpha$ at a regular grid of intervals of 2$mm$, 
with sphere radius $0.7$mm.  Each image has been 
normalised by dividing by the global norm of the average DT matrix. 
Dark red indicated low $\hat\alpha$ and light yellow indicates high $\hat\alpha$. 
There is a general increase in $\alpha$ when proceeding from left to right.}\label{FIG2}
\end{figure}

The results of the canine hearts analysis for the normalized data 
are displayed in Figures \ref{FIG1} and \ref{FIG2}. 
We see that $\hat\alpha$ is generally lower on the left of this region and higher on the right. 
Values are particularly high at the upper right in the view in Figure \ref{FIG2}.  
In practice one would wish to use the procedure to choose a single value of $\alpha$
suitable for the whole image, perhaps rounded to the nearest half for ease of explanation. 
Regarding the smoother estimates of Figure \ref{FIG1} values of $\alpha$ in the 
range $0.4-0.8$ would not
be unreasonable for this dataset, hence a suitable rounded choice would be 
$\alpha = 0.5$, corresponding to the matrix square root.

\begin{center}
{\bf INSERT FIGURES 1 AND 2 ABOUT HERE}
\end{center}

\begin{figure}[htbp]
\begin{center}
\includegraphics[width=8cm,angle=270]{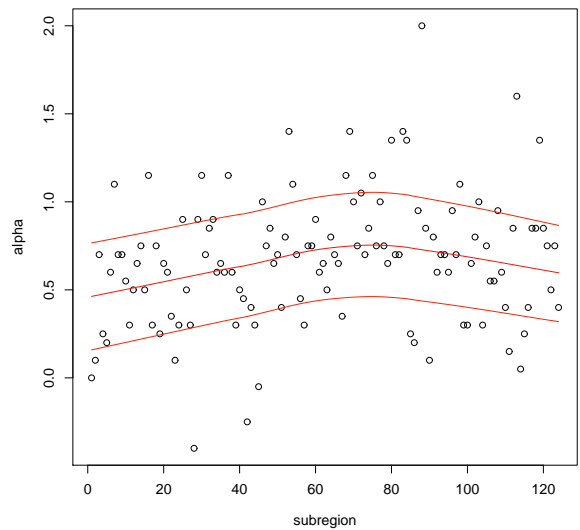}
\end{center}
\caption{The m.l.e. of $\alpha$ in each of the $124$ neighbourhoods for the unnormalized data. The location of the 
neighbourhoods proceeds from left of the mask (lower neighbourhood labels) to the right (higher numbered labels). 
A smoother estimate of the m.l.e. is given (middle line) together with smoothed upper and 
lower 95\% confidence limits for each neighbourhood.} \label{FIG3}
\end{figure}

\begin{figure}[htbp]
\begin{center}
\includegraphics[width=8cm]{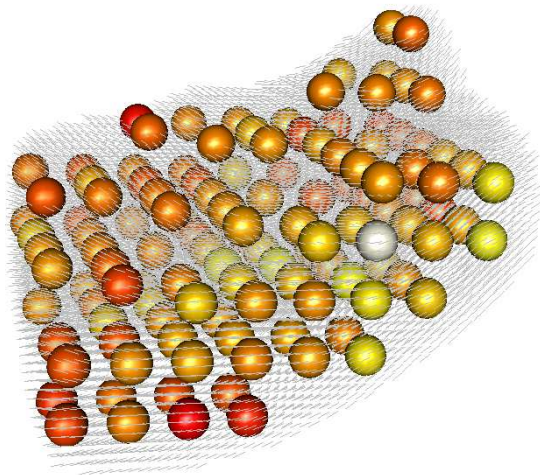}
\end{center}
\caption{The mask is shown with the principal axis of each DT for the first canine 
heart. The coloured spheres show the estimated $\alpha$ at a regular grid of intervals of 2$mm$, 
with sphere radius $0.7$mm, from the unnormalized data.
Dark red indicated low $\hat\alpha$ and light yellow indicates high $\hat\alpha$. }\label{FIG4}
\end{figure}

In Figures \ref{FIG3}-\ref{FIG4} we see the equivalent plots for the non-normalized data. 
Again values in the range $0.4-0.8$ seem reasonable, although there is no 
longer a  pattern of general 
increase from left to right. Rather the larger values are at the front/middle of the 
subregion. On average the estimates are a little lower than for the 
normalized data, but an overall rounded value of $\alpha = 0.5$ would be reasonable 
for these data as well.

\begin{center}
{\bf INSERT FIGURES 3 AND 4 ABOUT HERE}
\end{center}

Once a suitable value of $\alpha$ has been estimated then statistical analysis 
of the diffusion tensors can proceed. For example, we may wish to consider mean 
diffusion tensors in a  region using the Fr{\'e}chet mean, we might want to carry out interpolation in space 
between tensors or alternatively consider fibre tracking  using the metric 
corresponding to the estimated $\alpha$. \nocite{Alexander05} 

\section{Further work}
Rather than working with the symmetric power matrices, we could match two power matrices closer 
by post multiplying by an orthogonal matrix:
\begin{equation}
d_{P,\alpha}(S_1, S_2) =  
\mathop{\inf_{R \in O(m)}}   \frac{1}{\alpha} \|  S_1^{\alpha}  - S_2^{\alpha} R  \| . \label{dP}
\end{equation}
Note that $S^{\alpha}$ is symmetric, but $S^{\alpha} R$ is not symmetric in general. 
The motivation for carrying out this additional matching is that we can decompose 
$S$ as  
\begin{eqnarray}
S & = &  ([ S^{\alpha} RR^T S^{\alpha}] )^{1/(2 \alpha)} = ([(S^{\alpha}R)(S^{\alpha}R)^T] )^{1/(2\alpha)} , \nonumber
\end{eqnarray}
where $RR^T = R^TR = I_m$,  
and so the covariance matrix $S$ is invariant to the choice of $R$. The Euclidean 
power metric involves fixing $R=I$, with $S^{\alpha}$ symmetric. 
The Procrustes solution for matching $S_1^{\alpha}$ to $S_2^{\alpha}$ is 
$\hat R = U W^{\rm T}$, where $U,W$ are obtained from a singular value decomposition: 
$$
(S_2^\alpha)^{\rm T} S_1^{\alpha} =  W \Psi U^{\rm T}, \; \; U,W \in O(m) ,  
$$
and $\Psi$ is a diagonal matrix of singular values. 
The resulting metric is equivalent to Kendall's Riemannian size-and-shape metric
in the size-and-shape space of $m+1$ points in $m$ dimensions 
(see Dryden et al., 2009, when $\alpha = 1/2$). 
This non-Euclidean space has positive curvature, and is a cone with a warped product metric 
(Kendall, 1989; Dryden and Mardia, 1998). \nocite{Kendall89,Drydmard98} 
However, for practical use and in some simulation studies the Proctustes metric 
and the Euclidean power metric often perform similarly (see Dryden et al., 2009 in the 
$\alpha = 1/2$ case). Further properties and practical 
usage of the Procrustes power metric will be investigated in further work.

From the application point of view, more experimental data would be need
  to clearly establish what is the best metric for inter-subject
  comparison. It would also be very interesting to compare with the
  optimal metric for the intra-subject measurement noise. This last metric 
  could be estimated from repeated scan of the same subject.

\subsection*{Acknowledgements}

Funding for this work was provided in part by the Leverhulme Trust.

\bibliographystyle{apalike}
\bibliography{fullref}

\newpage

\begin{table}
\begin{center}
\begin{tabular}{|cc|c|}
\hline
$n_v$ & $n_s$ & Monte Carlo Coverage ($\%$)\\
\hline
$2$ & $4$ & $72.5$\\
$2$ & $5$ & $88.9$\\ %and 87.3
$3$ & $5$ & $93.8$\\
$4$ & $5$ & $95.0$\\
$5$ & $5$ & $94.4$\\
$5$ & $10$ & $96.1$\\
$10$&$10$ & $95.7$\\
$20$&$20$ & $95.5$\\
$20$&$30$ & $95.8$\\
\hline

\end{tabular}
\end{center}
\caption{Coverage from $1000$ Monte Carlo simulations for different values of
$n_v$ voxels and $n_s$ samples for the simulated data. The true coverage is $95\%$.}\label{TAB1}
\end{table}

\newpage



\end{document}